\documentclass[letterpaper,twocolumn,pra,aps,superscriptaddress,floatfix]{revtex4}
\usepackage{mathptmx}

\usepackage[latin9]{inputenc}
\usepackage{amsmath}
\usepackage{amssymb}
\usepackage{graphicx}
\usepackage{esint}
\usepackage{times}
\usepackage{array}

\usepackage[unicode=true,
 bookmarks=true,bookmarksnumbered=false,bookmarksopen=false,
 breaklinks=false,pdfborder={0 0 1},backref=false,colorlinks=true]
 {hyperref}
\hypersetup{linkcolor=magenta, urlcolor=blue, citecolor=blue, pdfstartview={FitH}, hyperfootnotes=false, unicode=true}

\makeatletter

\pdfpageheight\paperheight
\pdfpagewidth\paperwidth

\@ifundefined{textcolor}{}
{%
 \definecolor{BLACK}{gray}{0}
 \definecolor{WHITE}{gray}{1}
 \definecolor{RED}{rgb}{1,0,0}
 \definecolor{GREEN}{rgb}{0,1,0}
 \definecolor{BLUE}{rgb}{0,0,1}
 \definecolor{CYAN}{cmyk}{1,0,0,0}
 \definecolor{MAGENTA}{cmyk}{0,1,0,0}
 \definecolor{YELLOW}{cmyk}{0,0,1,0}
}

\usepackage{xcolor}\usepackage{soul}

\definecolor{blue}{rgb}{0,0,1}
\definecolor{red}{rgb}{1,0,0}
\definecolor{green}{rgb}{0,1,0}

\makeatother

\begin{document}

\title{Photon-Dressed Bloch-Siegert Shift in an Ultrastrongly Coupled Circuit Quantum Electrodynamical System}

\author{Shuai-Peng Wang}
\thanks{These authors contributed equally to this work.}
\affiliation{Quantum Physics and Quantum Information Division, Beijing Computational Science Research Center, Beijing 100193, China}
\affiliation{Interdisciplinary Center of Quantum Information, State Key Laboratory of Modern Optical Instrumentation, and Zhejiang Province Key Laboratory of Quantum Technology and Device, Department of Physics, Zhejiang University, Hangzhou 310027, China}

\author{Guo-Qiang Zhang}
\thanks{These authors contributed equally to this work.}
\affiliation{Interdisciplinary Center of Quantum Information, State Key Laboratory of Modern Optical Instrumentation, and Zhejiang Province Key Laboratory of Quantum Technology and Device, Department of Physics, Zhejiang University, Hangzhou 310027, China}

\author{Yimin Wang}
\affiliation{College of Communications Engineering, Army Engineering University, Nanjing 210007, China}

\author{Zhen Chen}
\affiliation{Quantum Physics and Quantum Information Division, Beijing Computational Science Research Center, Beijing 100193, China}
\affiliation{Interdisciplinary Center of Quantum Information, State Key Laboratory of Modern Optical Instrumentation, and Zhejiang Province Key Laboratory of Quantum Technology and Device, Department of Physics, Zhejiang University, Hangzhou 310027, China}

\author{Tiefu Li}
\thanks{litf@tsinghua.edu.cn}
\affiliation{Institute of Microelectronics, Tsinghua University, Beijing 100084, China}
\affiliation{Quantum Physics and Quantum Information Division, Beijing Computational Science Research Center, Beijing 100193, China}
\affiliation{Frontier Science Center for Quantum Information, Beijing 100084, China}
\affiliation{Beijing Academy of Quantum Information Sciences, Beijing 100193, China}

\author{J. S. Tsai}
\affiliation{Department of Physics, Tokyo University of Science, Kagurazaka, Shinjuku-ku, Tokyo 162-8601, Japan}
\affiliation{RIKEN Center for Emergent Matter Science (CEMS), 2-1 Hirosawa, Wako, Saitama 351-0198, Japan}

\author{Shi-Yao Zhu}
\affiliation{Interdisciplinary Center of Quantum Information, State Key Laboratory of Modern Optical Instrumentation, and Zhejiang Province Key Laboratory of Quantum Technology and Device, Department of Physics, Zhejiang University, Hangzhou 310027, China}

\author{J. Q. You}
\thanks{jqyou@zju.edu.cn}
\affiliation{Interdisciplinary Center of Quantum Information, State Key Laboratory of Modern Optical Instrumentation, and Zhejiang Province Key Laboratory of Quantum Technology and Device, Department of Physics, Zhejiang University, Hangzhou 310027, China}
\affiliation{Quantum Physics and Quantum Information Division, Beijing Computational Science Research Center, Beijing 100193, China}

\begin{abstract}
A cavity quantum electrodynamical (QED) system beyond the strong-coupling regime is expected to exhibit intriguing quantum phenomena. Here we report a direct measurement of the photon-dressed qubit transition frequencies up to four photons by harnessing the same type of state transitions in an ultrastrongly coupled circuit-QED system realized by inductively coupling a superconducting flux qubit to a coplanar-waveguide resonator. This demonstrates a convincing observation of the photon-dressed Bloch-Siegert shift in the ultrastrongly coupled quantum system. Moreover, our results show that the photon-dressed Bloch-Siegert shift becomes more pronounced as the photon number increases, which is a characteristic of the quantum Rabi model.
\end{abstract}

\maketitle

\section{Introduction}

The Jaynes-Cummings (JC) model has been widely used to describe the physics of a two-level quantum system (i.e., a qubit) coupled to the photons in a cavity when the qubit-photon interaction is in the regimes from weak to strong coupling~\cite{mabuchi2002cavity,raimond2001manipulating}. In these regimes, the rotating-wave approximation (RWA) is valid and the counter-rotating coupling terms can be ignored. However, when the coupling strength becomes even larger, the RWA breaks down and the counter-rotating terms play an increasingly important role. A single-mode quantum Rabi model, where both rotating- and counter-rotating coupling terms are included, has been harnessed to describe the cavity quantum electrodynamical (QED) system in an ultrastrong-coupling regime~\cite{kockum2019ultrastrong,forn2019ultrastrong, Devoret2007,Hanggi2009,Nataf-PRL2010, ashhab2010qubit, casanova2010deep, braak2011integrability,rossatto2017spectral,Ral-2017,Ral-2018}. In this quantum Rabi model, the whole excitation number of the system is no longer conserved, implying a variety of interesting quantum-optics phenomena, such as the virtual photon population in the ground state~\cite{liberato2009extracavity, garziano2013switching, lolli2015ancillary, stassi2013spontaneous, cirio2017amplified}, the generation of the entangled cat state~\cite{ashhab2010qubit, rossatto2017spectral} and the superradiant quantum phase transition in the classical oscillator limit~\cite{hwang2015quantum,liu2017universal,xie2014anisotropic}. These intriguing phenomena have attracted considerable attention in recent years, owing to the advancements in both theories and experiments. In addition to the analytic solution of the single-mode quantum Rabi model~\cite{braak2011integrability}, ultrastrong coupling was experimentally achieved in various cavity-QED systems~\cite{gunter2009sub,gambino2014exploring,bayer2017terahertz,li2018vacuum}.

As a circuit version of the cavity-QED system, the circuit-QED system consists of a superconducting qubit coupled to a coplanar-waveguide or lumped-element resonator~\cite{chiorescu2004coherent,wallraff2004strong}.  Benefiting from high degree of flexibility and controllability~\cite{nakamura1999coherent,You-Nature2011}, it provides an ideal platform to demonstrate various intriguing physical phenomena in the ultrastrong-~\cite{niemczyk2010circuit, forn2010observation, forn2016broken, bosman2017multi, forn2017continuum, chen2017single, magazzu2018probing} and even deep-strong-coupling regimes~\cite{yoshihara2017superconducting,yoshihara2018inversion}. Also, several protocols based on the circuit-QED system for quantum information processing were proposed by utilizing the unique features in the ultrastrong-coupling regime~\cite{nataf2011protected, romero2012ultrafast, wang2016holonomic, stassi2018long}. Here we report a direct observation of the photon-dressed Bloch-Siegert shift in an ultrastrongly coupled flux-qubit-based circuit-QED system. The Bloch-Siegert shift is a signature that the RWA breaks down and can be used as a measure for characterizing the deviation of the quantum Rabi model from the JC model~\cite{li2018vacuum,forn2010observation}. In the previous experiment of the ultrastrongly coupled circuit-QED system, vacuum Bloch-Siegert shift was observed~\cite{forn2010observation, forn2016broken} by studying the qubit-state-dressed photon transition that reduces to $|g,0\rangle\rightarrow|g,1\rangle$ in the absence of the qubit-resonator coupling, where $g$ denotes the ground state of the qubit and $0,1$ denote zero and one photons in the resonator mode, respectively. As shown below, {\it only} the vacuum Bloch-Siegert shift can be measured even for the qubit-state-dressed photon transitions related to $|g,N\rangle\rightarrow|g,N+1\rangle$. In Ref.~\cite{yoshihara2018inversion}, photon-dressed qubit transition frequencies up to two photons were investigated in a deep-strongly-coupled circuit QED system, but they are {\it indirectly} derived via various types of state transitions in the system and only the values at the optimal point of the flux qubit were obtained. In our experiment, by designing a different setup, i.e., a multi-mode system with a relatively weakly-coupled resonator mode as a probe to extract information from the ultrastrongly coupled system, we can {\it directly} measure the photon-dressed qubit transition frequencies just using the {\it same} type of state transitions. This makes it more convenient to measure the Bloch-Siegert shifts involving multiple photons. Indeed, as observed in our results, it is experimentally realizable to measure the photon-dressed Bloch-Siegert shifts up to four photons in a larger range of the flux bias around the optimal point of the flux qubit.

\section{The ultrastrongly coupled circuit-QED system}

\begin{figure}[tb]
\includegraphics[width=3.4in, keepaspectratio=ture]{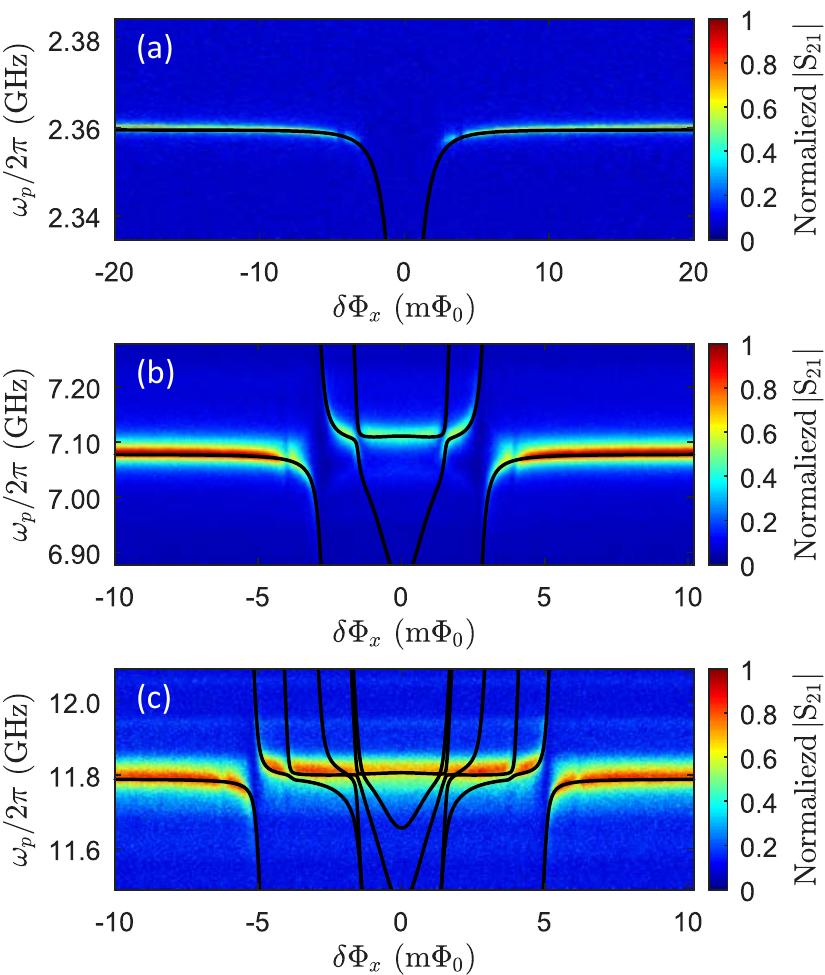}
\caption{Transmission spectra versus the external flux bias $\delta \Phi$ and probe frequency $\omega_p$ around (a)~$\lambda /2$, (b)~$3\lambda /2$, and (c)~$5\lambda /2$ modes of the resonator. The black solid curves are numerical fits to the measured spectra using the transition energies calculated from the multi-mode quantum Rabi Hamiltonian (\ref{Hamil}). In (c), many curves intersect near $\delta \Phi_x\equiv\Phi_x-\Phi_0/2= \pm 1.5~{\mathrm m}\Phi_0$, forming several small splittings, but these small splittings become indistinguishable in the data due to the larger damping rate of the $5\lambda /2$ mode.}
\label{fig:fig1}
\end{figure}

The quantum circuit that we used comprises a superconducting flux qubit galvanically coupled to a $\lambda/2$-type superconducting coplanar-waveguide resonator~\cite{niemczyk2010circuit,chen2017single} (see Appendix~\ref{AppA}). The flux qubit consists of four Josephson junctions with three identical larger junctions and a smaller junction reduced by a factor of 0.6 in area. Both experimental~\cite{Bertet2005} and theoretical~\cite{Qiu2016} studies show that the four-junction flux qubit behaves similar to the three-junction flux qubit~\cite{Orlando-1999}. The design of the quantum circuit is also similar to that in Ref.~\cite{chen2017single} but has different parameters. In the basis of the persistent-current states $\{|\circlearrowleft\rangle,\;|\circlearrowright\rangle\}$,
the Hamiltonian of the flux qubit can be written as
$H_q=\epsilon \tau_z+\Delta \tau_x$.
Here $\tau_{z}$ and $\tau_{x}$ are Pauli operators, and $\Delta $ is the hopping amplitude between the two persistent-current states. The offset energy induced by a flux bias is
$\epsilon =2I_p(\Phi_x-\Phi_0/2)$,
where $I_p$ is the maximal persistent current, $\Phi_x$ is the externally applied flux threading through the loop of the flux qubit, and $\Phi_0$ is the magnetic flux quantum.

To realize an {\it ultrastrong} coupling between the flux qubit and the resonator, part of the central conductor of the niobium superconducting coplanar-waveguide resonator is replaced by an aluminum strip (length 34.8 $\mu$m, width 800 nm, and thickness 60 nm) which is also shared by the flux qubit as part of the loop. Compared with the $LC$ resonator~\cite{yoshihara2017superconducting,yoshihara2018inversion}, the $\lambda/2$-type coplanar-waveguide resonator has  multiple modes and its Hamiltonian is written as (we set $\hbar=1$)
$H_r=\sum_m \omega_{r,m}\left(a^{\dagger}_m a_m+\frac{1}{2}\right)$,
where $a^{\dagger}_m$ ($a_m$) is the creation (annihilation) operator of the $m$th resonator mode (i.e., the $m\lambda /2$ mode) and $\omega_{r,m}$ is the corresponding resonance frequency. In the design of our circuit-QED system, the loop of the flux qubit is located around the {\it common} current antinode of the odd-number modes, so the flux qubit is mainly coupled to the odd-number modes of the coplanar-waveguide resonator and its coupling to the even-number modes of the resonator are {\it negligibly} small~\cite{chen2017single}. The frequencies of the lowest three odd-number modes (i.e., $m=1,3,5$) are $\omega_{r,1}=2\pi\times 2.360$~GHz, $\omega_{r,3}=2\pi\times 7.078$~GHz, and $\omega_{r,5}=2\pi\times 11.789$~GHz, as determined by a transmission measurement. The magnetic-dipolar interaction Hamiltonian is
$H_{\rm int}=\sum_m g_m(a^{\dagger}_m+a_m)\tau_z$,
where $g_m=MI_pI_{r,m}$ is the inductive-coupling strength between the flux qubit and the $m$th resonator mode, with $M$ being the mutual inductance between the flux qubit and the resonator, and $I_{r,m}$ the vacuum central-conductor current of the $m$th resonator mode. When the basis of the flux qubit is converted to the eigenbasis of the qubit $\{|g\rangle,|e\rangle\}$, the full Hamiltonian of the circuit-QED system, $H=H_q+H_r+H_{\rm int}$, can be written as a generalized multi-mode quantum Rabi model,
\begin{eqnarray}
H&=&\frac{1}{2}\omega_q\sigma_z+\sum_m \omega_{r,m}\left(a^{\dagger}_m a_m+\frac{1}{2}\right) \nonumber\\
&&+\sum_m g_m[\cos(\theta)\sigma_z-\sin(\theta)\sigma_x](a^{\dagger}_m+a_m),
\label{Hamil}
\end{eqnarray}
where $\omega_q=\sqrt{\epsilon^2+\Delta^2}$ is the transition frequency of the qubit, $\sigma_z$ and $\sigma_x$ are Pauli operators related to the eigenbasis of the flux qubit, and $\tan(\theta)=\Delta/\epsilon$.

The whole circuit-QED system was mounted inside a dilution refrigerator cooled down to $\sim 30$~mK. At such a low temperature, both the flux qubit and the resonator were nearly in their ground states. To measure the transmission spectrum of the coplanar-waveguide resonator embedding the flux qubit, we applied a weak probe signal $\omega_p$ in the proximity of a resonance frequency of the resonator and measured the transmission amplitude using a vector network analyzer. The probe signal was kept weak enough to maintain that the average photon number in the resonator was less than one. Figure~\ref{fig:fig1} shows the transmission spectra around the $\lambda /2$, $3\lambda /2$, and $5\lambda /2$ modes, respectively. We can derive the coupling strengths $g_n$ ($n=1,3,5$) by fitting the transition energies calculated using the full Hamiltonian $H$ to the measured spectra. These coupling strengths are obtained as $g_1=2\pi\times 265$~MHz, $g_3=2\pi\times 459$~MHz, and $g_5=2\pi\times 592$~MHz. Normalized to the corresponding resonance frequencies, we have the ratios $g_1/\omega_{r,1}=11.2\%$, $g_3/\omega_{r,3}=6.5\%$, and $g_5/\omega_{r,5}=5.0\%$, implying that the coupling of the flux qubit to the $\lambda /2$ mode is clearly in the {\it ultrastrong-coupling} regime. From the fitting, we also obtain the maximal persistent current $I_p=360$~nA and the hopping amplitude $\Delta =2\pi\times 3.198$~GHz.

\begin{figure}[tb]
\includegraphics[width=3.4in, keepaspectratio=ture]{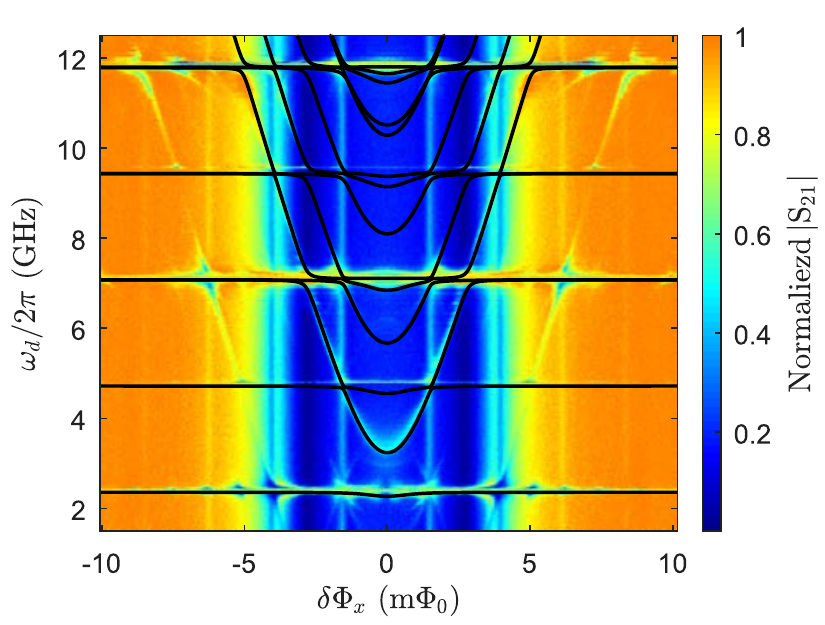}
\caption{State-transition spectrum versus the external flux bias $\delta \Phi_x$ and drive frequency $\omega_d$.
The frequency of the probe tone is fixed at $7.078$~GHz, in resonance with the effective resonant frequency of the $3\lambda/2$ mode at $\delta \Phi_x=10~{\mathrm m}\Phi_0$.
The numerical transition frequencies (black solid curves) between the ground state and the excited eigenstates of the circuit-QED system are calculated using the multi-mode Rabi Hamiltonian (\ref{Hamil}), with the parameters obtained from the fittings in Fig.~\ref{fig:fig1}.}
\label{fig:fig2}
\end{figure}

\section{Photon-dressed Bloch-Siegert shift}

In order to demonstrate the spectrum of the circuit-QED system in a wider range of the frequency, we also performed the two-tone spectroscopy of the coplanar-waveguide resonator embedding the flux qubit by applying a weak probe tone with its frequency {\it fixed} at the resonance frequency of the $3\lambda/2$ mode. As shown above, the interaction between the $3\lambda/2$ mode and the flux qubit is not in the ultrastrong-coupling regime. The effect of the probe field on the system can be reduced by choosing the probe-tone frequency at the $3\lambda/2$ mode,  instead of the $\lambda/2$ mode. Meanwhile, the frequency of a strong drive tone was scanned to produce the allowed transitions between the ground state and the excited eigenstates of the circuit-QED system. These allowed transitions versus the external flux bias correspond to the brightest curves shown in Fig.~\ref{fig:fig2}, where the drive tone was applied via the same input port of the resonator as the probe tone. The solid curves in Fig.~\ref{fig:fig2} are the numerical results for the transition frequencies between the ground state and the excited eigenstates calculated using the Hamiltonian $H$ in Eq.~(\ref{Hamil}), with $m=1,3,5$, by varying the external flux bias. It is clear that these transition frequencies match well the brightest curves observed in the experiment. In addition to these brightest curves, there are extra less bright curves (i.e., those not corresponding to the solid curves in Fig.~\ref{fig:fig2}) which actually correspond to the sideband transitions discussed in Ref.~\cite{chen2017single}.

\begin{figure}[tb]
\includegraphics*[width=3.5in, keepaspectratio=ture]{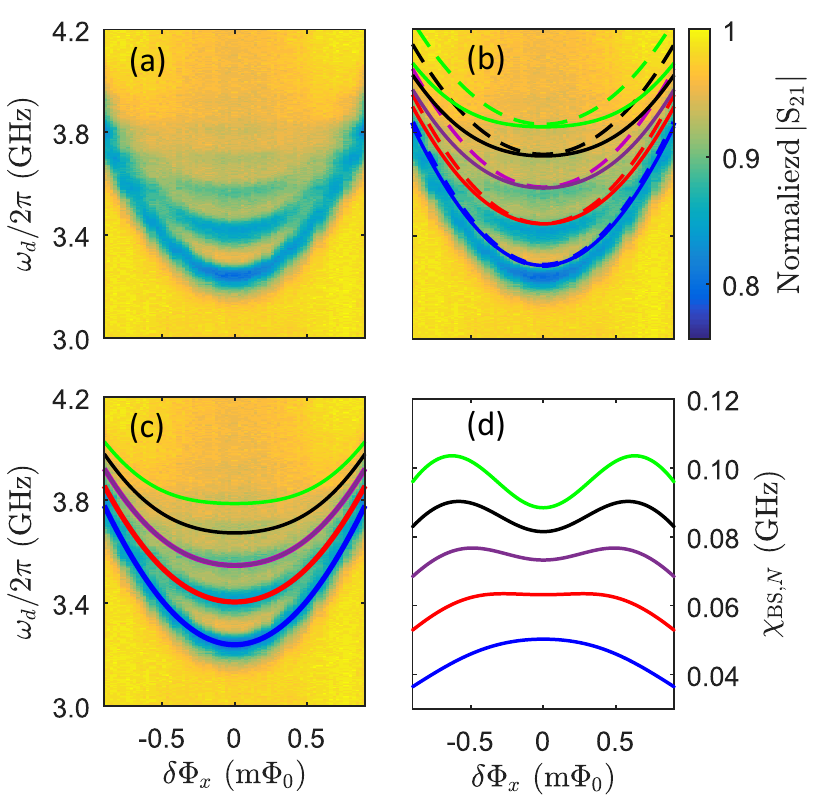}
\caption{(a)~A zoom-in view of the photon-dressed qubit transition spectrum near $\omega_d/2\pi=3.5$~GHz versus the external flux bias around the optimal point of the flux qubit $\delta \Phi_x=0$. The frequency of the probe tone is fixed at $7.106$~GHz, in resonance with the effective resonant frequency of the $3\lambda/2$ mode at $\delta {\mathrm{\Phi }}_x=0$. Compared to the case in Fig.~\ref{fig:fig2}, the drive power is tuned much lower to reduce the power-induced broadening of the qubit linewidth, so as to have the discrete transition curves well resolved. (b)~The dashed (solid) curves from bottom to top are the analytic (numerical) results for the photon-dressed qubit transition frequencies with $N=0,1,2,3$ and 4, which are calculated, respectively, using Eq.~(\ref{OmegaBS}) and the Rabi Hamiltonian (\ref{Hamil}) including only the $\lambda/2$ mode of the resonator. (c) The solid curves are the numerical results corresponding to those in (b), but the $\lambda/2$, $3\lambda/2$ and $5\lambda/2$ modes of the resonator are included in the Rabi Hamiltonian (\ref{Hamil}).
(d) Photon-dressed Bloch-Siegert shift $\chi_{{\rm BS},N}$ versus the external flux bias $\delta \Phi_x$. The curves from bottom to top correspond to $N=0,1,2,3$ and 4, respectively, and are obtained from the solid curves in (c) by subtracting the corresponding
qubit transition frequencies calculated using the Hamiltonian  (\ref{Hamil}) but excluding the counter-rotating coupling terms.}
\label{fig:fig3}
\end{figure}

The lowest parabolic curve around $\omega_d/2\pi=3.5$~GHz corresponds to the state transition of the circuit-QED system that reduces to the transition between $|g,0\rangle$ and $|e,0\rangle$ in the absence of the qubit-resonator coupling, where the zero corresponds to the vacuum state of the lowest $\lambda/2$ mode. Vacuum states are also assured around the frequency $\omega_d/2\pi=3.5$~GHz for other resonator modes because their frequencies are much higher. Here we denote the frequency of this transition as $\omega_{q,0}$. In fact, other state transitions having frequencies $\omega_{q,N}$ can occur for the ultrastrongly coupled circuit-QED system, which are close to $\omega_{q,0}$ but correspond to the transitions between $|g,N\rangle$ and $|e,N\rangle$ in the absence of the qubit-resonator coupling, where $N=1,2,3,\dots$ denote the nonzero number of photons in the $\lambda/2$ mode of the resonator. We find that these transitions can be resolved when reducing the drive-tone intensity, as shown in Fig.~\ref{fig:fig3}(a). These photon-dressed qubit transitions were not discovered in ultrastrongly coupled circuit-QED
systems~\cite{niemczyk2010circuit,forn2010observation,chen2017single} but now become resolved in the present experiment, owing to the improved coherence of the flux qubit (which gives rise to a reduced linewidth of the qubit) and the harnessed experimental setup and measurement method.

It is worth pointing out that in our setup, the lowest $\lambda/2$ mode of the resonator is ultrastrongly coupled to the flux qubit, but no signal is applied on this mode to obtain the results in Fig.~\ref{fig:fig3}(a). Thus, the photon numbers involved in the observed photon-dressed qubit transitions in Fig.~\ref{fig:fig3}(a) are related to the background thermal photons in the $\lambda/2$ mode, which is also confirmed by the standard Bose-Einstein distribution that monotonically decreases when increasing the photon number. The intrinsic linewidth of the flux qubit is expected to be about few tens of MHz, inferred from the observed qubit-transition linewidths in Fig.~\ref{fig:fig3}(a). For the bare $\lambda/2$ mode of the resonator, it has a linewidth of 0.8 MHz, as can be determined from the transmission spectrum in Fig.~\ref{fig:fig1}(a). Due to the strong hybridization with the flux qubit, the effective linewidth of the $\lambda/2$ mode near the optimal flux bias point is much broadened. This allows more thermal photons to occur, as more frequency components of the background thermal noise can enter the resonator. As estimated from the experimental results, the average thermal photon number $\langle\hat n_1\rangle$ is about 3 in our setup when the flux qubit is near the optimal flux bias point (Appendix \ref{AppB}).

\begin{figure}[tb]
\includegraphics*[width=3.5in, keepaspectratio=ture]{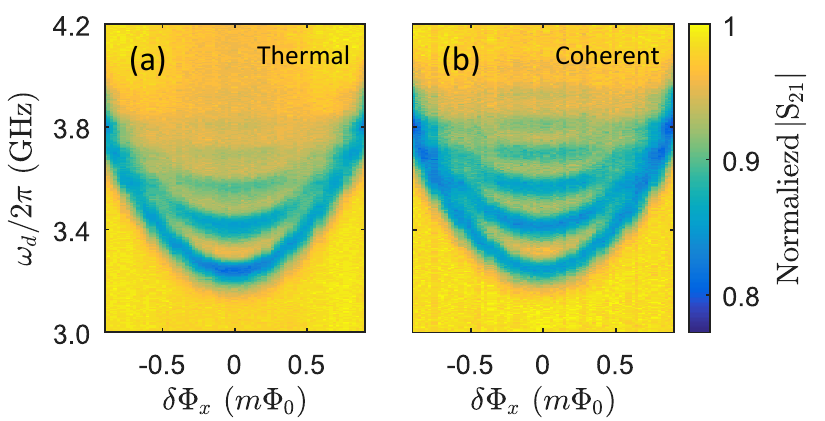}
\caption{(a)~Thermal and (b) coherent-state spectra of the photon-dressed qubit transitions for the $\lambda/2$ mode of the resonator with average photon number $\langle\hat n_1\rangle\sim 3$. The thermal spectrum in (a) is monotonic and consistent with the standard Bose-Einstein distribution, $P(n)={\langle\hat n\rangle}^n/{\langle\hat n+1\rangle}^{n+1}$, while the coherent-state spectrum in (b) is nonmonotonic and qualitatively consistent with the Poisson distribution, $P(n)=e^{-\langle\hat n\rangle}{\langle\hat n\rangle}^n/n!$.}
\label{fig:fig4}
\end{figure}

In Fig.~\ref{fig:fig4}, we further compare the thermal spectrum with another spectrum obtained by applying an additional weak coherent drive tone on the lowest $\lambda/2$ mode. With the weak coherent drive tone applied on the $\lambda/2$ mode, the spectrum clearly shows the coherent-state spectrum which is nonmonotonic and qualitatively consistent with the Poisson distribution. In such a case, we cannot apply a strong drive tone on the $\lambda/2$ mode because this resonator mode is ultrastrongly coupled to the flux qubit. Otherwise, the strong drive tone will yield the linewidths of the photon-dressed qubit transitions too broad to resolve the fine structure in Fig.~\ref{fig:fig4}(b), cf. Appendix \ref{AppC}. This can be explained as follows. A strong drive tone will heat the resonator much and equivalently raise the environmental temperature of the flux qubit. Then, it reduces the quantum coherence of the flux qubit by broadening the linewidth of the qubit transition (i.e., increasing the relaxation rate of the qubit).

To understand the photon-dressed Bloch-Siegert shift, it is useful to first only consider the ultrastong coupling with the $\lambda/2$ mode. When $\omega_q+\omega_{r,1}\gg g_1$, by keeping the leading terms (Appendix~\ref{AppD}), the generalized single-mode Rabi Hamiltonian can be converted to a standard Bloch-Siegert Hamiltonian~\cite{Note1},
\begin{eqnarray}
H_{\rm BS}&=&\frac{\omega_q}{2}\sigma_z+\omega_{r,1}\left(\hat n_1+\frac{1}{2}\right)+\omega_{\text{BS}}\left[\sigma_z\left(\hat n_1+\frac{1}{2}\right)-\frac{1}{2}\right] \nonumber\\
&&-g_1\sin(\theta)(a_1^{\dag}\sigma_-+ {a_1}\sigma_+),
\label{HamilBS}
\end{eqnarray}
where $\hat n_1= a_1^{\dag}a_1$, and $\omega_{\rm BS} \equiv g_1^2 \sin^2(\theta)/(\omega_q+\omega_{r,1})$ is the vacuum Bloch-Siegert shift. In the limit of $\omega_{\rm BS}\to0$, the normal JC model is recoverd.
We consider the case of $\delta_{N}\equiv\omega_q-\omega_{r,1}+2N\omega_{\rm BS}>0$ because $\omega_q>\omega_{r,1}$ in our experiment. The photon-dressed qubit transition frequency can be derived by diagonalizing the Bloch-Siegert Hamiltonian (Appendix~\ref{AppD}),
\begin{equation}
\omega^{({\rm Rabi})}_{q,N}=\omega_{r,1}+\frac{1}{2}\left(\sqrt{\delta^2_{N+1}+4g^2_{1,N+1}}
+\sqrt{\delta^2_{N}+4g^2_{1,N}}\right),
\label{OmegaBS}
\end{equation}
where $g_{1,N}=-g_1\sin(\theta)\sqrt{N}$. When $\omega_{{\rm BS}}\to0$, $\omega^{({\rm Rabi})}_{q,N}\to \omega^{({\rm JC})}_{q,N}$.
We can define the $N$-photon-dressed Bloch-Siegert shift of the qubit transition frequency as
$\chi_{{\rm BS},N}=\omega^{({\rm Rabi})}_{q,N}-\omega^{({\rm JC})}_{q,N}$.
In the dispersive regime of $\omega_q-\omega_{r,1}\gg 2|g_{1,N}|$, $\omega^{({\rm Rabi})}_{q,N}\approx \omega^{({\rm JC})}_{q,N} +(2N+1)\omega_{\rm BS}$, giving rise to a simple expression for the photon-dressed Bloch-Siegert shift $\chi_{{\rm BS},N}\approx (2N+1)\omega_{\rm BS}$.
In our experiment, $\omega_q-\omega_{r,1}\sim 3g_{1}$. We cannot use this simple expression of $\chi_{{\rm BS},N}$ to fit the experimental results, because the dispersive condition is not fully satisfied.

We fit the experimental results in Fig.~\ref{fig:fig3}(a) using {\it both} the photon-dressed qubit transition frequency given in Eq.~(\ref{OmegaBS}) {\it and} the numerical results obtained from the single-mode Rabi Hamiltonian, i.e., Eq.~(\ref{Hamil}) with only the $\lambda/2$ mode included. It is clear that the analytic expression in Eq.~(\ref{OmegaBS}) (dashed curves) can qualitatively reproduce the qubit transition curves, except for a small overall upward frequency shift in each case of $N$ as can be seen in Fig.~\ref{fig:fig3}(b). Near the optimal flux bias point $\delta {\mathrm{\Phi }}_x=0$, these analytical results are also close to the corresponding numerical results (solid curves) obtained from the single-mode Rabi Hamiltonian, but they deviate when away from the optimal flux bias point, because the longitudinal coupling terms proportional to $\cos(\theta)$, which is neglected in the reduced Bloch-Siegert Hamiltonian (\ref{HamilBS}), becomes nonzero when $\delta {\mathrm{\Phi }}_x\neq 0$.
In the experiment around $\omega_d/2\pi=3.5$~GHz, $|\omega_q-\omega_{r,1}|\sim 3g_{1}$,  $|\omega_q-\omega_{r,3}|\sim 8g_{3}$, and $|\omega_q-\omega_{r,5}|\sim 14g_{5}$, where $g_{1}$, $g_{3}$ and $g_{5}$ were measured above to be $g_1=2\pi\times 265$~MHz, $g_3=2\pi\times 459$~MHz, and $g_5=2\pi\times 592$~MHz. The $3\lambda/2$ mode of the resonator is expected to produce an appreciable effect on the state transitions of the circuit-QED system at $\omega_d/2\pi\sim 3.5$~GHz, while the $5\lambda/2$ mode gives rise to much less appreciable effect on these state transitions, owing to the large frequency detuning (see Fig.~\ref{fig:fig8} in Appendix {\ref{AppE}).
Figure~\ref{fig:fig3}(c) presents the numerical results calculated using the multi-mode Rabi Hamiltonian in Eq.~(\ref{Hamil}) by including the $\lambda/2$, $3\lambda/2$ and $5\lambda/2$ modes of the resonator. Now, the overall upward frequency shifts are corrected and the obtained numerical results are in an excellent agreement with the experimental observations, further confirming that the ultrastrongly coupled circuit-QED system can be well described by a multi-mode Rabi Hamiltonian. In Fig.~\ref{fig:fig3}(d), we also present the photon-dressed Bloch-Siegert shifts $\chi_{{\rm BS},N}=\omega^{({\rm Rabi})}_{q,N}-\omega^{({\rm JC})}_{q,N}$, where $\omega_{q,N}^{\rm (Rabi)}$ correspond to the solid curves in Fig.~\ref{fig:fig3}(c)
and $\omega_{q,N}^{\rm (JC)}$ are the corresponding results calculated by excluding the counter-rotating coupling terms in the multi-mode Rabi Hamiltonian (\ref{Hamil}). The increase of the photon-dressed Bloch-Siegert shift with the photon number is clearly demonstrated.

\section{Discussions and Conclusions}

In Ref.~\cite{forn2010observation}, vacuum Bloch-Siegert shift was observed via the qubit-state-dressed photon transition that reduces to the transition $|g,0\rangle\rightarrow|g,1\rangle$ in the absence of the qubit-resonator coupling. When $\omega_q-\omega_{r,1}+2(N+1)\omega_{\rm BS}<0$, as similar to the case in Ref.~\cite{forn2010observation}, the photon transition frequency, related to $|g,N\rangle\rightarrow|g,N+1\rangle$  in the absence of the qubit-resonator coupling, can be obtained from the Bloch-Siegert Hamiltonian in Eq.~(\ref{HamilBS}) as (Appendix~\ref{AppF})
\begin{equation}\label{eq4}
\omega^{({\rm Rabi})}_{N,g}=\omega_{r,1}+\frac{1}{2}\left(\sqrt{\delta^2_{N+1}+4g^2_{1,N+1}}
-\sqrt{\delta^2_{N}+4g^2_{1,N}}\right).
\end{equation}
In the dispersive regime of $\omega_{r,1}-\omega_q\gg 2|g_{1,N}|$, $\omega^{({\rm Rabi})}_{N,g}\approx\omega^{({\rm JC})}_{N,g} -\omega_{\rm BS}$, only giving the vacuum Bloch-Siegert shift $\chi_{{\rm BS},N}\approx -\omega_{\rm BS}$, {\it irrespective} of the photon number $N$. Therefore, qubit-state-dressed photon transitions cannot be used to measure the Bloch-Siegert shifts involving multiple photons. In our experiment, we design a different setup, i.e., a multi-mode system with a relatively weakly-coupled resonator mode as a probe to extract information from the ultrastrongly coupled system. In the previous studies~\cite{niemczyk2010circuit, forn2010observation, forn2016broken}, the system involves either single mode or multi-modes of the resonator but all modes are too strongly coupled with the qubit. In these cases, because of the high hybridization, probing the resonator mode will result in significant disturbance on the qubit and the qubit linewidth is usually too broad to resolve the fine structure in Fig.~\ref{fig:fig3} and Fig.~\ref{fig:fig4}. However, with our setup, by just using the {\it same} type of photon-dressed qubit transitions, we can {\it directly} resolve the photon-dressed Bloch-Siegert shifts up to four photons in a large range of the flux bias around the optimal point of the flux qubit. Also, this is in sharp contrast to the more complicated case in Ref.~\cite{yoshihara2018inversion}, where the photon-dressed qubit transition frequencies up to two photons were observed indirectly at {\it only} the optimal point of the flux qubit by harnessing several {\it different} types of state transitions in a deep-strongly-coupled quantum system.

In conclusion, we have observed the photon-dressed Bloch-Siegert shifts up to four photons in an ultrastrongly coupled circuit-QED system. This platform is expected to explore more quantum-optics phenomena when the coherence of the flux qubit is further improved, e.g., using a capacitively shunted flux qubit~\cite{CSFQ-1,CSFQ-2,CSFQ-3,NTT-2019}, in which the large shunt capacitance can considerably reduce the sensitivity of the flux qubit to the charge noise.
Also, we may use the present experimental setup with some modifications, e.g., inserting Josephson junctions in the shared part of the qubit loop to greatly enhance the coupling strength~\cite{yoshihara2017superconducting, forn2017continuum, magazzu2018probing, yoshihara2018inversion}, to measure a more complete energy spectrum of a deep-strongly-coupled circuit-QED system and possibly detect the virtual photons in the ground state of the deep-strongly-coupled quantum system.

\begin{acknowledgments}
This work is supported by the National Key Research and Development Program of China (Grant No.~2016YFA0301200),  Science Challenge Project (Grant No.~TZ2018003), the National Natural Science Foundation of China (Grants No.~11934010, No.~U1801661, and No.~U1930402), and BAQIS Research Program (Grant No.~Y18G27). J.S.T. is partially supported by the JST CREST, NEDO IoT, and Japanese cabinet office ImPACT.
\end{acknowledgments}

\appendix

\begin{figure*}[tb]
\includegraphics*[width=7.0in, keepaspectratio=ture]{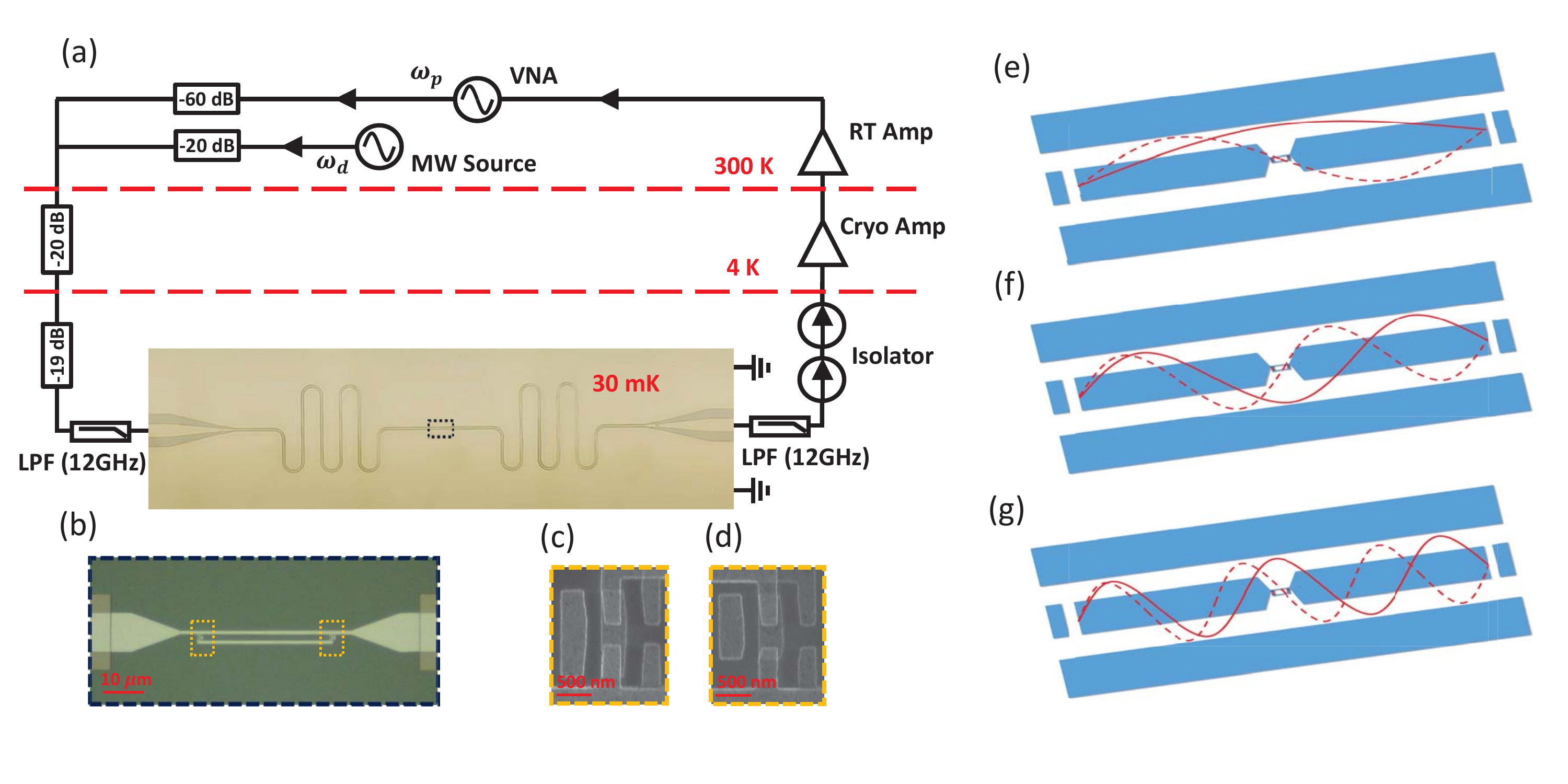}
\caption{(a)~Schematic of the experimental setup. (b)~A zoom-in view of the optical image denoted by the blue rectangular box in (a), where the four-junction flux qubit is galvanically coupled to the coplanar-waveguide resonator via a shared arm between the qubit and the resonator's central line. (c) and (d)~Zoom-in views of the scan electron microscopy (SEM) images denoted by the left and right yellow rectangular boxes in (b), respectively. (e)~Current distributions in the central line of the resonator for the $\lambda/2$ (solid curve) and $\lambda$ (dashed curve) modes of the resonator. (f)~Current distributions in the central line of the resonator for the $3\lambda/2$ (solid curve) and $2\lambda$ (dashed curve) modes of the resonator.
(g)~Current distributions in the central line of the resonator for the $5\lambda/2$ (solid curve) and $3\lambda$ (dashed curve) modes of the resonator.}
\label{fig:fig5}
\end{figure*}

\section{Experimental Setup}\label{AppA}

Transmission spectra of the coplanar-waveguide resonator at the frequency $\omega_p$ of the probe tone are measured using a vector network analyser (VNA). Another microwave signal at frequency $\omega_d$ is further harnessed for the two-tone spectroscopy measurements. The input signals are attenuated and filtered at various temperature stages before finally reaching the coplanar-waveguide resonator in which the flux qubit is embedded [Fig.~\ref{fig:fig5}(a)]. Also, two isolators and a low-pass filter (LPF) are used to protect the sample from the amplifier's noise. In our design of the sample, the flux qubit is galvanically connected to the coplanar-waveguide resonator via a shared arm between the qubit and the resonator's central line [Fig.~\ref{fig:fig5}(b)]. As part of the qubit's loop, this shared arm is 34.8~$\mu$m long and 800~nm wide. The flux qubit has four Josephson junctions in the loop [Figs.~\ref{fig:fig5}(c) and \ref{fig:fig5}(d)], with three identical larger junctions and a smaller junction reduced by a factor of $0.6$ in area. The loop of the flux qubit is fabricated at the center of the coplanar-waveguide resonator and the modes $m\lambda/2$ of the resonator, with $m=1$ to 6, are here considered. Around the flux qubit, the current distribution in the central line of the coplanar-waveguide resonator reaches an antinode for each of the odd-number modes $\lambda/2$, $3\lambda/2$ and $5\lambda/2$ [see the solid curves in Figs.~\ref{fig:fig5}(e)-\ref{fig:fig5}(g)], and an ultrastrong or strong coupling is achieved between the qubit and the corresponding odd-number mode of the resonator (see the main text). However, for each of the even-number modes $\lambda$, $2\lambda$ and $3\lambda$, the current distribution in the central line of the coplanar-waveguide resonator approaches zero around the flux qubit [see the dashed curves in Figs.~\ref{fig:fig5}(e)-\ref{fig:fig5}(g)]. This gives rise to a weak coupling between the qubit and the corresponding even-number mode of the resonator.

\section{Estimation of the background thermal photon number in the $\lambda/2$ mode}\label{AppB}

Figure~\ref{fig:fig6} shows the transmission decay $1-|S_{21}|$ versus the photon number $n$, as extracted from the results in Fig.~\ref{fig:fig3}(a) at the optimal point of the flux qubit. We can estimate the background thermal photon number in the $\lambda/2$ mode of the resonator by fitting the transmission decay at different photon number $n$ using the formula ${\langle\hat n\rangle}^n/{\langle\hat n+1\rangle}^{n+1}$, which corresponds to the Bose-Einstein distribution. From the fitting in Fig.~\ref{fig:fig6}, we obtain $\langle\hat n_1\rangle\sim 3$ for the $\lambda/2$ mode.

\begin{figure}[hbt!]
\includegraphics*[width=3.4in, keepaspectratio=ture]{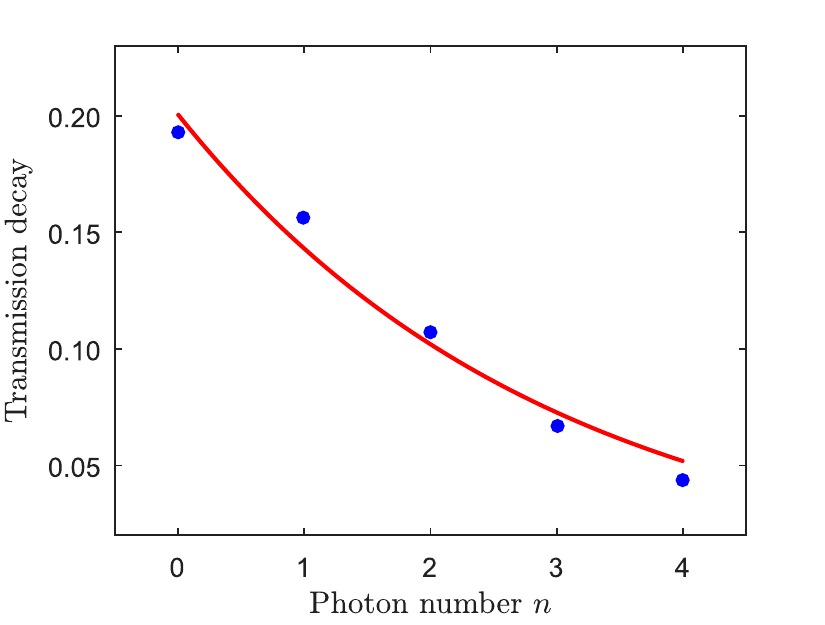}
\caption{Fitting the transmission decay $1-|S_{21}|$ at different photon number $n$ with a Bose-Einstein distribution ${\langle\hat n\rangle}^n/{\langle\hat n+1\rangle}^{n+1}$. The experimental data (blue dots) are extracted from Fig.~\ref{fig:fig3}(a) at the optimal flux bias point.}
\label{fig:fig6}
\end{figure}

\section{Photon-dressed qubit transition spectra with increasing drive power applied to the $\lambda/2$ mode}\label{AppC}

As the drive power applied to the $\lambda/2$ mode of the resonator gradually increases, the photon-dressed qubit transition spectrum changes from a thermal spectrum [cf. Fig.~\ref{fig:fig4}(a)] (corresponding to zero or an extremely weak drive power) to a coherent-state spectrum [cf. Fig.~\ref{fig:fig4}(b)] (corresponding to a weak drive power). To show the linewidth broadening of the photon-dressed qubit transitions induced by the drive power, we implemented the measurement for a wider range of the drive power. Figure~\ref{fig:fig7} presents the transmission decay $1-|S_{21}|$ versus the frequency $\omega_d$ of the drive tone at the optimal flux bias point by applying various drive powers to the $\lambda/2$ mode of the resonator. It can be seen that when the drive tone is sufficiently strong, the linewidths of the photon-dressed qubit transitions become too broad to resolve the fine structure observed at a weaker drive power (see the top curve in Fig.~\ref{fig:fig7}).

\begin{figure}[hbt!]
\includegraphics*[width=3.4in, keepaspectratio=false]{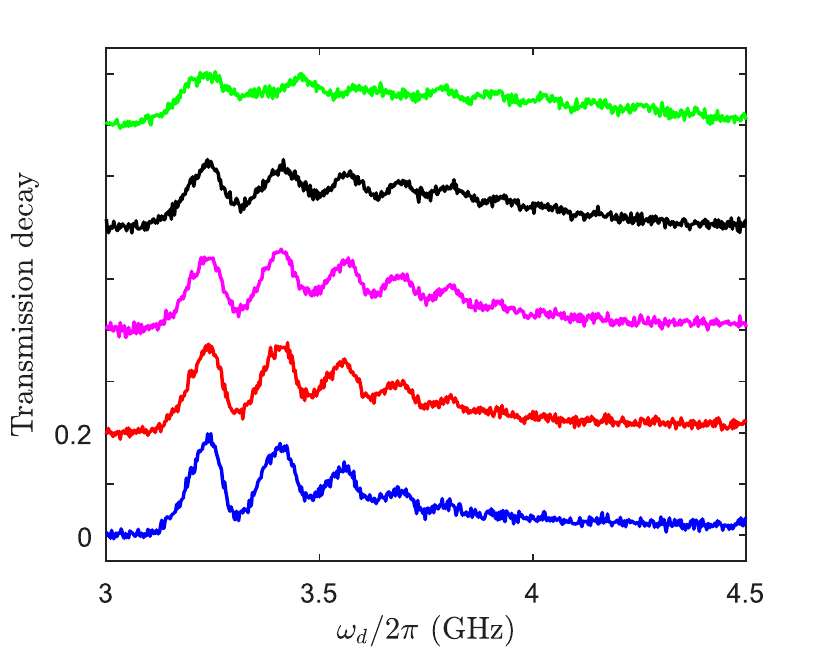}
\caption{Photon-dressed qubit transition spectra with increasing drive powers applied to the $\lambda/2$ mode of the resonator, where the flux bias is at the optimal point of the flux qubit. From bottom to top curves, the drive powers applied at the device input port are $-145$~dBm, $-140$~dBm, $-135$~dBm, $-130$~dBm, $-125$~dBm, respectively. For clarity, the transition spectra are offset vertically.}
\label{fig:fig7}
\end{figure}

\section{Derivation of the Bloch-Siegert Hamiltonian in Eq.~(\ref{HamilBS})}\label{AppD}

When only considering the $\lambda/2$ mode of the resonator, the multi-mode quantum Rabi Hamiltonian in Eq.~(\ref{Hamil}) reduces to a generalized single-mode quantum Rabi model,
\begin{eqnarray}
H\!&\!=\!&\!\frac{1}{2}\omega_q\sigma_z+\omega_{r,1}\left(a^{\dagger}_1 a_1+\frac{1}{2}\right) \nonumber\\ &&\!+g_1[\cos(\theta)\sigma_z-\sin(\theta)\sigma_x](a^{\dagger}_1+a_1).
\label{S1}
\end{eqnarray}
For simplicity, we denote $\omega_{r,1}$ as $\omega_r$, and $a_1$ ($a^{\dagger}_1$) as $a$ ($a^{\dagger}$).  Also, we rewrite the Hamiltonian $H$ as $H=H_0+H_I$, with
\begin{eqnarray}
H_{0}\!&\!=\!&\!\frac{1}{2}\omega_{q}\sigma_{z}+\omega_{r}\bigg(a^{\dag}a+\frac{1}{2}\bigg), \\
H_{I}\!&\!=\!&\!-g\sin(\theta)(a^{\dag}+a)(\sigma_{+}+\sigma_{-})+g\cos(\theta)(a^{\dag}+a)\sigma_{z},\nonumber
\end{eqnarray}
where $\sigma_{\pm}=(\sigma_{x} \pm i\sigma_{y})/2$ are the rising and lowering operators of the qubit.

To convert $H$ to a Hamiltonian with only rotating coupling terms, we can introduce a canonical
transformation~\cite{Klimov2009,forn2010observation} $\exp(-S)$, with
$S=\gamma(a^{\dag}\sigma_{+}-a\sigma_{-})$, where $\gamma=-g\sin(\theta)/(\omega_{q}+\omega_{r})$.
In our experimental setup, $\omega_q=2\pi\times 3.198$~GHz at the optimal point $\delta\Phi_x=0$ of the flux qubit,
while $g=2\pi\times 265$~MHz and $\omega_r=2\pi\times 2.360$~GHz. Thus, $|\gamma|\sim 0.05\ll 1$. Applying this canonical transformation to the Hamiltonian $H$, we have
\begin{eqnarray}
H_{\rm eff}\!&\!=\!&\!e^{S}He^{-S} \nonumber\\
      &\!=\!&\!H+[S,H]+\frac{1}{2}[S,[S,H]]+\cdots \\
      &\!=\!&\!H_{0}+(H_{I}+[S,H_{0}])+\bigg([S,H_{I}]+\frac{1}{2}[S,[S,H_{0}]]\bigg)+\cdots.\nonumber
\end{eqnarray}
The first-order terms are
\begin{eqnarray}
H_{I}+[S,H_{0}]\!&\!=\!&\!-g\sin(\theta)(a^{\dag}\sigma_{-}+a\sigma_{+}) \nonumber\\
&&\!+g\cos(\theta)(a^{\dag}+a)\sigma_{z},
\end{eqnarray}
and the second-order terms are
\begin{eqnarray}
[S,H_{I}]\!&\!=\!&\! -2\gamma g\sin(\theta)\left[\left(a^{\dag}a+\frac{1}{2}\right)\sigma_{z}-\frac{1}{2}\right] \nonumber\\
            &&\!-\gamma g\sin(\theta)(a^{\dag}a^{\dag}+aa)\sigma_{z} \nonumber\\
         &&\!-2\gamma g\cos(\theta)(a^{\dag}a^{\dag}\sigma_{+}+aa\sigma_{-})\nonumber\\
            &&\!-\gamma g \cos(\theta)(2a^{\dag}a+1)(\sigma_{+}+\sigma_{-}); \nonumber\\
\frac{1}{2}[S,[S,H_{0}]]\!&\!=\!&\!\frac{1}{2}\left[\gamma(a^{\dag}\sigma_{+}-a\sigma_{-}),
                                                    g\sin(\theta)(a^{\dag}\sigma_{+}+a\sigma_{-})\right]\nonumber\\
                        \!&\!=\!&\!\gamma g\sin(\theta) \left[\left(a^{\dag}a+\frac{1}{2}\right)\sigma_{z}-\frac{1}{2}\right].
\end{eqnarray}
Up to the second order, the effective Hamiltonian is
\begin{eqnarray}
H_{\rm eff}\!&\!=\!&\!H_{0}+(H_{I}+[S,H_{0}])+\bigg([S,H_{I}]+\frac{1}{2}[S,[S,H_{0}]]\bigg)\nonumber\\
      &\!=\!&\! {H}_{\rm BS}+H_{l}-\gamma g\sin(\theta)(a^{\dag}a^{\dag}+aa)\sigma_{z},
\label{Heff}
\end{eqnarray}
where $H_{\rm BS}$ is the Bloch-Siegert Hamiltonian,
\begin{eqnarray}
H_{\rm BS}\!&\!=\!&\!\frac{1}{2}\omega_{q}\sigma_{z}+\omega_{r}\bigg(a^{\dag}a+\frac{1}{2}\bigg)
                  +\omega_{\rm BS}\bigg[\bigg(a^{\dag}a+\frac{1}{2}\bigg)\sigma_{z}-\frac{1}{2}\bigg] \nonumber\\
                  &&\! -g\sin(\theta)(a^{\dag}\sigma_{-}+a\sigma_{+}),
\label{Hbs}
\end{eqnarray}
with
$\omega_{\rm BS}=g^{2}\sin^{2}(\theta)/(\omega_{q}+\omega_{r})$
being the vacuum Bloch-Siegert shift, and
$H_l$ is a resulting Hamiltonian due to the longitudinal terms,
\begin{eqnarray}
H_l \!&\!=\!&\! g\cos(\theta)(a^{\dag}+a)\sigma_{z}-2\gamma g\cos(\theta)(a^{\dag}a^{\dag}\sigma_{+}+aa\sigma_{-}) \nonumber\\
           &&\! -\gamma g \cos(\theta)(2a^{\dag}a+1)(\sigma_{+}+\sigma_{-}).
\end{eqnarray}
For the standard quantum Rabi model without longitudinal terms, it corresponds to the case with the flux qubit
at the optimal point $\delta\Phi_x=0$, i.e., $\theta=\pi/2$ in Eq.~(\ref{S1}). In fact, the longitudinal terms in $H_l$ are counter-rotating terms, which are less important than the rotating terms in Eq.~(\ref{Hbs}) and are also proportional to $\cos(\theta)$. In our work, we focus on the region close to the optimal point $\theta=\pi/2$, where $\cos(\theta)\sim 0$.  Thus, in comparison with the rotating terms in Eq.~(\ref{Hbs}), these counter-rotating terms can be neglected when $\theta$ is close to $\pi/2$.
The third term in Eq.~(\ref{Heff}), $-\gamma g\sin(\theta)(a^{\dag}a^{\dag}+aa)\sigma_{z}$, is also a counter-rotating term. As given above, $|\gamma|\sim 0.05\ll 1$, so $|\gamma|g\ll g$. Compared with the rotating coupling terms in Eq.~(\ref{Hbs}), the term $-\gamma g\sin(\theta)(a^{\dag}a^{\dag}+aa)\sigma_{z}$ can also be neglected in our experimental setup. Therefore, the effective Hamiltonian can be reduced to the Bloch-Siegert Hamiltonian in Eq.~(\ref{Hbs}), i.e., $H_{\rm eff}\approx H_{\rm BS}$, when tuning $\theta$ close to $\theta=\pi/2$.

As in Ref.~\cite{Klimov2009}, another canonical transformation can be harnessed to remove the third term $-\gamma g\sin(\theta)(a^{\dag}a^{\dag}+aa)\sigma_{z}$ in Eq.~(\ref{Heff}), but it yields higher-order rotating coupling terms in the Bloch-Siegert Hamiltonian~\cite{forn2010observation}. These higher-order rotating terms have coupling strength
\begin{equation}
g\sin(\theta)\frac{\omega_{\rm BS}}{\omega_q+\omega_r}=\gamma^2 g\sin(\theta)\sim 0.0025g,
\end{equation}
which, compared to $g$, is negligible here. After neglecting these higher-order rotating terms, the effective Hamiltonian given in Ref.~\cite{forn2010observation} is also reduced to the Bloch-Siegert Hamiltonian in Eq.~(\ref{Hbs}).

\section{Photon-Dressed Qubit Transition Frequency}\label{AppE}

In the basis $\{|g,N+1\rangle,|e,N\rangle\}$, where $g$ and $e$ denote, respectively, the ground and excited states of the flux qubit, and $N$ denotes the Fock state of the resonator mode with $N$ photons, the Bloch-Siegert Hamiltonian $H_{\rm BS}$ is represented as a block diagonal matrix and each block is a $2\times 2$ matrix
\begin{widetext}
\begin{equation}
H_{{\rm BS},N+1}=
\left(
  \begin{array}{cc}
    -\frac{1}{2}\omega_{q}+(N+\frac{3}{2})\omega_{r}-(N+2)\omega_{\rm BS} & g_{N+1} \\
   g_{N+1} & \frac{1}{2}\omega_{q}+(N+\frac{1}{2})\omega_{r}+N\omega_{\rm BS}\\
  \end{array}
\right),
\end{equation}
\end{widetext}
where
$g_{N}=-g\sin(\theta)\sqrt{N}$.
This matrix has two eigenvalues
\begin{equation}\label{eigen}
\omega_{N+1,\pm}=\left(N+1\right)\omega_{r}-\omega_{\rm BS}
                     \pm \frac{1}{2}\sqrt{\delta_{N+1}^{2}+4g_{N+1}^{2}},
\end{equation}
with
$\delta_{N}=(\omega_{q}-\omega_{r})+2N\omega_{\rm BS}$.
In the limit $\omega_{\rm BS}\rightarrow 0$, the eigenvalues in Eq.~(\ref{eigen}) reduce to the results of the corresponding Jaynes-Cummings (JC) model.

(i)~{\it The case of $\omega_q-\omega_r+2N\omega_{\rm BS}>0$}. We define
\begin{eqnarray}
\omega_{N,e} \!&\! \equiv \!&\! \omega_{N+1,+}=\left(N+1\right)\omega_{r}-\omega_{\rm BS}
                     + \frac{1}{2}\sqrt{\delta_{N+1}^{2}+4g_{N+1}^{2}},\nonumber\\
\omega_{N,g} \!&\! \equiv \!&\! \omega_{N,-}=N\omega_{r}-\omega_{\rm BS}
                     - \frac{1}{2}\sqrt{\delta_{N}^{2}+4g_{N}^{2}}.
\end{eqnarray}
The ground-state energy of the system is $\omega_{0,g}\equiv \omega_{0,-}=-\frac{1}{2}(\omega_{q}-\omega_{r})-\omega_{\rm BS}$. In the absence of the qubit-resonator coupling, i.e., $g=0$, $\omega_{N,e}$ and $\omega_{N,g}$ reduce to
$\omega_{N,e}\equiv \omega_{N+1,+}=(N+\frac{1}{2})\omega_{r}+\frac{1}{2}\omega_q$ and
$\omega_{N,g}\equiv \omega_{N,-}=(N+\frac{1}{2})\omega_{r}-\frac{1}{2}\omega_q$,
which are the energies of the states $|e,N\rangle$ and $|g,N\rangle$, respectively. Thus,
\begin{eqnarray}\label{f-QT}
\omega_{q,N}^{({\rm Rabi})} \!&\!\equiv\!&\! \omega_{N,e}- \omega_{N,g} \nonumber\\
&\!=\!&\! \omega_{r}+ \frac{1}{2}\left(\sqrt{\delta_{N+1}^{2}+4g_{N+1}^{2}}
                                               +\sqrt{\delta_{N}^{2}+4g_{N}^{2}}\right)~~~~~~~~
\end{eqnarray}
is the frequency of the photon-dressed qubit transition that reduces to $|g,N\rangle\rightarrow|e,N\rangle$ in the absence of the qubit-resonator coupling. Equation~(\ref{f-QT}) is just Eq.~(\ref{OmegaBS}) in the main text
as $\omega_r$ and $g_N$ are replaced by $\omega_{1,r}$ and $g_{1,N}$, respectively.
The photon-dressed qubit transition frequency $\omega_{q,N}^{({\rm JC})}$ of the corresponding JC model is given by
$\omega_{q,N}^{({\rm Rabi})}$ in the limit of $\omega_{\rm BS}\rightarrow 0$.

In the dispersive regime with $\omega_q-\omega_r\gg 2|g_{N+1}|$,
\begin{eqnarray}
\frac{1}{2}\sqrt{\delta_{N+1}^{2}+4g_{N+1}^{2}}\!&\!\approx\!&\!\frac{1}{2}\left[(\omega_q-\omega_r)+2(N+1)\omega_{\rm BS}\right]
\nonumber\\
&&\!+\frac{(N+1)g^2\sin^2(\theta)}{\omega_q-\omega_r},
\end{eqnarray}
so we have
\begin{eqnarray}\label{f-Rabi}
\omega_{q,N}^{({\rm Rabi})}\!&\!\approx \!&\! \omega_q+(2N+1)\omega_{\rm BS}+\frac{(2N+1)g^2\sin^2(\theta)}{\omega_q-\omega_r} \nonumber\\
&\!=\!&\! \omega_{q,N}^{({\rm JC})}+ (2N+1)\omega_{\rm BS}.
\end{eqnarray}
The photon-dressed Bloch-Siegert shift is
\begin{equation}\label{BS-shift} \chi_{{\rm BS},N}\equiv\omega_{q,N}^{({\rm Rabi})}-\omega_{q,N}^{({\rm JC})}\approx (2N+1)\omega_{\rm BS},
\end{equation}
which is equally spaced by $2\omega_{\rm BS}$.

\begin{figure}[tb]
\includegraphics*[width=3.6in, keepaspectratio=ture]{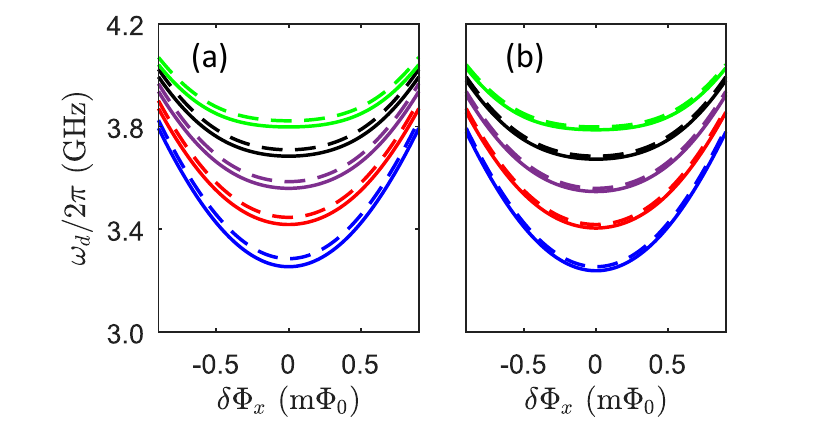}
\caption{Photon-dressed qubit transition frequencies with $N=0,1,2,3$ and 4 (curves from bottom to top), as obtained using the multi-mode Rabi Hamiltonian (\ref{Hamil}) in the main text by including different modes of the resonator. (a)~The dashed curves are the results when only the $\lambda/2$ mode is included, while the solid curves are the results when both $\lambda/2$ and $3\lambda/2$ modes are included. (b)~The dashed curves are the results when both $\lambda/2$ and $3\lambda/2$ modes are included, while the solid curves are the results when $\lambda/2$, $3\lambda/2$ and $5\lambda/2$ modes are included.}
\label{fig:fig8}
\end{figure}

In Fig.~\ref{fig:fig3}(b), we have compared the photon-dressed qubit transition frequencies calculated using the Hamiltonian
in Eq.~(\ref{S1}) with those obtained using the analytical result in Eq.~(\ref{f-QT}).
As expected, they are close to each other near the optimal point $\delta\Phi_x=0$ of the flux qubit, but deviate away from
$\delta\Phi_x=0$. The numerical results obtained using Eq.~(\ref{S1}) can qualitatively reproduce the qubit
transition curves observed in our experiment, except for a small overall upward frequency shift in each case of $N$.
In addition to the $\lambda/2$ mode of the resonator, when the $3\lambda/2$ and $5\lambda/2$ modes are included,
the numerical results match well the experimentally observed qubit transition curves [cf. Fig.~\ref{fig:fig3}(c)].
Here we show the numerical results for the photon-dressed qubit transition frequency obtained using the multi-mode Rabi Hamiltonian (\ref{Hamil}) in the main text by including different modes of the resonator. When both $\lambda/2$ and $3\lambda/2$ modes are included, the numerical results have appreciable difference from those obtained by including only the $\lambda/2$ mode [Fig.~\ref{fig:fig8}(a)]. When further including the $5\lambda/2$ mode, the numerical results appear slightly different from those obtained by including only the $\lambda/2$ and $3\lambda/2$ modes [Fig.~\ref{fig:fig8}(b)].
This clearly reveals that the resonator mode $5\lambda/2$ has much less effect on the photon-dressed qubit transitions than the resonator
mode $3\lambda/2$, owing to the larger freqency detuning at the experimental condition. For the resonator modes
$m\lambda/2$ with $m>5$, their effects on the photon-dressed qubit transitions should be negligibly small and can thus be ignored,
because the corresponding frequency detunings become even larger.

(ii)~{\it The case of $\omega_q-\omega_r+2(N+1)\omega_{\rm BS}<0$}. Similarly, we can define
\begin{eqnarray}
\omega_{N,e} \!&\! \equiv \!&\! \omega_{N+1,-}=\left(N+1\right)\omega_{r}-\omega_{\rm BS}
                     - \frac{1}{2}\sqrt{\delta_{N+1}^{2}+4g_{N+1}^{2}},\nonumber\\
\omega_{N,g} \!&\! \equiv \!&\! \omega_{N,+}=N\omega_{r}-\omega_{\rm BS}
                     + \frac{1}{2}\sqrt{\delta_{N}^{2}+4g_{N}^{2}}.
\end{eqnarray}
The ground-state energy of the system is now given by
$\omega_{0,g}\equiv \omega_{0,+}=-\frac{1}{2}(\omega_{q}-\omega_{r})-\omega_{\rm BS}$.
In the absence of the qubit-resonator coupling, $\omega_{N,e}$ and $\omega_{N,g}$ also reduce to
$\omega_{N,e}\equiv \omega_{N+1,-}=(N+\frac{1}{2})\omega_{r}+\frac{1}{2}\omega_q$ and
$\omega_{N,g}\equiv \omega_{N,+}=(N+\frac{1}{2})\omega_{r}-\frac{1}{2}\omega_q$,
namely, the energies of the states $|e,N\rangle$ and $|g,N\rangle$, respectively. The photon-dressed qubit transition frequency is
\begin{eqnarray}
\omega_{q,N}^{({\rm Rabi})} \!&\!\equiv\!&\! \omega_{N,e}- \omega_{N,g} \nonumber\\
&\!=\!&\! \omega_{r}- \frac{1}{2}\left(\sqrt{\delta_{N+1}^{2}+4g_{N+1}^{2}}
                                               +\sqrt{\delta_{N}^{2}+4g_{N}^{2}}\right).~~~~~~~~
\end{eqnarray}
In the dispersive regime with $\omega_r-\omega_q\gg 2|g_{N+1}|$, we can obtain the same result as
in Eq.~(\ref{f-Rabi}) and the photon-dressed Bloch-Siegert shift $\chi_{{\rm BS},N}$ is also given by Eq.~(\ref{BS-shift}).

\section{Qubit-State-Dressed Photon Transition Frequency}\label{AppF}

(i)~{\it The case of $\omega_q-\omega_r+2(N+1)\omega_{\rm BS}<0$}.  We define
\begin{eqnarray}
\omega_{N+1,g} \!&\! \equiv \!&\! \omega_{N+1,+}=\left(N+1\right)\omega_{r}-\omega_{\rm BS}
                     + \frac{1}{2}\sqrt{\delta_{N+1}^{2}+4g_{N+1}^{2}},\nonumber\\
\omega_{N,g} \!&\! \equiv \!&\! \omega_{N,+}=N\omega_{r}-\omega_{\rm BS}
                     + \frac{1}{2}\sqrt{\delta_{N}^{2}+4g_{N}^{2}}.
\end{eqnarray}
The ground-state energy of the system is $\omega_{0,g}\equiv \omega_{0,+}=-\frac{1}{2}(\omega_{q}-\omega_{r})-\omega_{\rm BS}$.
In the absence of the qubit-resonator coupling, $\omega_{N+1,g}$ and $\omega_{N,g}$ reduce to
$\omega_{N+1,g} \equiv \omega_{N+1,+}=(N+\frac{3}{2})\omega_r-\frac{1}{2}\omega_q$ and
$\omega_{N,g}  \equiv  \omega_{N,+}=(N+\frac{1}{2})\omega_r-\frac{1}{2}\omega_q$, which are the energies of
the states $|g,N+1\rangle$ and $|g,N\rangle$, respectively. Thus,
\begin{eqnarray}\label{f-PT}
\omega_{N,g}^{({\rm Rabi})} \!&\!\equiv\!&\! \omega_{N+1,g}-\omega_{N,g} \nonumber\\
&\!=\!&\! \omega_{r}+ \frac{1}{2}\left(\sqrt{\delta_{N+1}^{2}+4g_{N+1}^{2}}    -\sqrt{\delta_{N}^{2}+4g_{N}^{2}}\right)~~~~~~~~
\end{eqnarray}
is the frequency of the qubit-state-dressed photon transition that reduces to $|g,N\rangle\rightarrow|g,N+1\rangle$
in the absence of the qubit-resonator coupling. Equation~(\ref{f-PT}) is just Eq.~(\ref{eq4}) in the main text
as $\omega_r$ and $g_N$ are replaced by $\omega_{1,r}$ and $g_{1,N}$, respectively.

In the dispersive regime with $\omega_r-\omega_q\gg 2|g_{N+1}|$,
\begin{eqnarray}
\frac{1}{2}\sqrt{\delta_{N+1}^{2}+4g_{N+1}^{2}}\!&\!\approx\!&\! -\frac{1}{2}\left[(\omega_q-\omega_r)+2(N+1)\omega_{\rm BS}\right]                      \nonumber\\
&&\!+\frac{(N+1)g^2\sin^2(\theta)}{\omega_r-\omega_q}.
\end{eqnarray}
Then, we have
\begin{eqnarray}\label{f-Rabi-1}
\omega_{N,g}^{({\rm Rabi})}\!&\!\approx \!&\! \omega_r-\omega_{\rm BS}+\frac{g^2\sin^2(\theta)}{\omega_r-\omega_q} \nonumber\\
&\!=\!&\! \omega_{N,g}^{({\rm JC})}-\omega_{\rm BS}.
\end{eqnarray}                                                                                                                   The Bloch-Siegert shift is
\begin{equation}\label{BS-shift-1}
\chi_{{\rm BS},N}\equiv\omega_{N,g}^{({\rm Rabi})}-\omega_{N,g}^{({\rm JC})}\approx -\omega_{\rm BS}.
\end{equation}
In sharp constrast to the photon-dressed Bloch-Siegert shift in Eq.~(\ref{BS-shift}), this Bloch-Siegert shift is irrespective of
the photon number $N$.

Also, we can define
\begin{eqnarray}
\omega_{N,e} \!&\! \equiv \!&\! \omega_{N+1,-}=\left(N+1\right)\omega_{r}-\omega_{\rm BS}
                     - \frac{1}{2}\sqrt{\delta_{N+1}^{2}+4g_{N+1}^{2}},\nonumber\\
\omega_{N-1,e} \!&\! \equiv \!&\! \omega_{N,-}=N\omega_{r}-\omega_{\rm BS}
                     - \frac{1}{2}\sqrt{\delta_{N}^{2}+4g_{N}^{2}}.
\end{eqnarray}
In the absence of the qubit-resonator coupling, $\omega_{N,e}$ and $\omega_{N-1,e}$ reduce to
$\omega_{N,e} \equiv \omega_{N+1,-}=(N+\frac{1}{2})\omega_r+\frac{1}{2}\omega_q$ and
$\omega_{N-1,e}  \equiv  \omega_{N,-}=[(N-1)+\frac{1}{2}]\omega_r+\frac{1}{2}\omega_q$, which are the energies of
the states $|e,N\rangle$ and $|e,N-1\rangle$, respectively. Then,
\begin{eqnarray}\label{}
\omega_{N,e}^{({\rm Rabi})} \!&\!\equiv\!&\! \omega_{N,e}-\omega_{N-1,e} \nonumber\\
\!&=\!&\!\omega_{r}- \frac{1}{2}\left(\sqrt{\delta_{N+1}^{2}+4g_{N+1}^{2}}-\sqrt{\delta_{N}^{2}+4g_{N}^{2}}\right)~~~~~~~~
\end{eqnarray}
is the frequency of the qubit-state-dressed photon transition that reduces to $|e,N-1\rangle\rightarrow|e,N\rangle$
in the absence of the qubit-resonator coupling.
In the dispersive regime with $\omega_r-\omega_q\gg 2|g_{N+1}|$, we have
\begin{eqnarray}\label{f-Rabi-2}
\omega_{N,e}^{({\rm Rabi})} \!&\!\approx \!&\! \omega_r+\omega_{\rm BS}-\frac{g^2\sin^2(\theta)}{\omega_r-\omega_q} \nonumber\\
&\!=\!&\! \omega_{N,e}^{({\rm JC})}+\omega_{\rm BS},
\end{eqnarray}
and the Bloch-Siegert shift is
\begin{equation}\label{BS-shift-2}
\chi_{{\rm BS},N}\equiv\omega_{N,e}^{({\rm Rabi})}-\omega_{N,e}^{({\rm JC})}\approx \omega_{\rm BS},
\end{equation}
which is also irrespective of the photon number $N$.

(ii)~{\it The case of $\omega_q-\omega_r+2N\omega_{\rm BS}>0$}. Similarly, we can define
\begin{eqnarray}
\omega_{N+1,g} \!&\! \equiv \!&\! \omega_{N+1,-}=\left(N+1\right)\omega_{r}-\omega_{\rm BS}
                     - \frac{1}{2}\sqrt{\delta_{N+1}^{2}+4g_{N+1}^{2}},\nonumber\\
\omega_{N,g} \!&\! \equiv \!&\! \omega_{N,-}=N\omega_{r}-\omega_{\rm BS}
                     - \frac{1}{2}\sqrt{\delta_{N}^{2}+4g_{N}^{2}}.
\end{eqnarray}
In this case, the ground-state energy of the system is
$\omega_{0,g}\equiv \omega_{0,-}=-\frac{1}{2}(\omega_{q}-\omega_{r})-\omega_{\rm BS}$.
The qubit-state-dressed photon transition frequency is
\begin{eqnarray}
\omega_{N,g}^{({\rm Rabi})} \!&\!\equiv\!&\! \omega_{N+1,g}-\omega_{N,g} \nonumber\\
&\!=\!&\!\omega_{r}- \frac{1}{2}\left(\sqrt{\delta_{N+1}^{2}+4g_{N+1}^{2}}
                                               -\sqrt{\delta_{N}^{2}+4g_{N}^{2}}\right).~~~~~~~~~~
\end{eqnarray}
In the dispersive regime with $\omega_q-\omega_r\gg 2|g_{N+1}|$, we can obtain the same result as
in Eq.~(\ref{f-Rabi-1}) and the Bloch-Siegert shift $\chi_{{\rm BS},N}$ is also given by Eq.~(\ref{BS-shift-1}).

Also, we can define
\begin{eqnarray}
\omega_{N,e} \!&\! \equiv \!&\! \omega_{N+1,+}=\left(N+1\right)\omega_{r}-\omega_{\rm BS}
                     + \frac{1}{2}\sqrt{\delta_{N+1}^{2}+4g_{N+1}^{2}},\nonumber\\
\omega_{N-1,e} \!&\! \equiv \!&\! \omega_{N,+}=N\omega_{r}-\omega_{\rm BS}
                     + \frac{1}{2}\sqrt{\delta_{N}^{2}+4g_{N}^{2}}.
\end{eqnarray}
The corresponding qubit-state-dressed photon transition frequency is
\begin{eqnarray}\label{}
\omega_{N,e}^{({\rm Rabi})} \!&\!\equiv\!&\! \omega_{N,e}-\omega_{N-1,e} \nonumber\\
&\!=\!&\!\omega_{r}+ \frac{1}{2}\left(\sqrt{\delta_{N+1}^{2}+4g_{N+1}^{2}}-\sqrt{\delta_{N}^{2}+4g_{N}^{2}}\right).~~~~~~~~~~
\end{eqnarray}
In the dispersive regime with $\omega_q-\omega_r\gg 2|g_{N+1}|$,
we have the same result as in Eq.~(\ref{f-Rabi-2}) and
the Bloch-Siegert shift $\chi_{{\rm BS},N}$ is also given by Eq.~(\ref{BS-shift-2}).

\end{document}